\documentclass[aps,prb,reprint,groupedaddress,showkeys]{revtex4-2}

\usepackage{graphicx}
\usepackage{longtable}
\usepackage{calc}
\usepackage{amsmath}
\usepackage{multirow}
\usepackage{xcolor}
\usepackage{float}

\usepackage[T1]{fontenc}
\usepackage{subfigure}
\usepackage{booktabs}
\usepackage{enumitem}
\newlist{examples}{enumerate}{1}

\begin{document}

\title{Electronic and Vibrational Properties of Layered Boron Nitride Polymorphs}

\author{Priyanka Mishra and Nevill Gonzalez Szwacki}
\affiliation{Faculty of Physics, University of Warsaw, Pasteura 5, PL-02093 Warsaw, Poland}

\begin{abstract}
We present a comprehensive first-principles investigation of the structural, electronic, and vibrational properties of four layered boron nitride (BN) polymorphs--AA-stacked ($e$-BN), AA$^\prime$-stacked ($h$-BN), ABC-stacked ($r$-BN), and AB-stacked ($b$-BN). Using density functional theory and density functional perturbation theory with and without van der Waals (vdW) corrections, we quantify the impact of interlayer dispersion on lattice parameters, electronic band gaps, phonon frequencies, and infrared and Raman intensities. Our results demonstrate that vdW interactions are essential for reproducing experimental lattice constants and stabilizing interlayer phonon modes. The vibrational spectra exhibit distinct stacking-dependent features, enabling clear differentiation among polytypes. Notably, $b$-BN displays a direct band gap, while $r$-BN shows enhanced IR and Raman activity due to LO-TO splitting and symmetry breaking. These findings underscore the critical role of interlayer interactions in determining the physical properties of $sp^2$-bonded BN and offer insight into the experimental identification and functionalization of BN polytypes for electronic and photonic applications.
\end{abstract}

\maketitle


\section{Introduction}

Hexagonal boron nitride ($h$-BN) is a prototypical layered material with a honeycomb lattice structure analogous to graphite, composed of strongly bonded in-plane B--N atoms ($sp^2$ hybridization) and weakly interacting interlayer planes. This structural anisotropy imparts $h$-BN with a wide band gap, exceptional thermal and chemical stability, and a rich vibrational spectrum, positioning it as a vital material in diverse applications including deep ultraviolet optoelectronics, two-dimensional (2D) heterostructures, and quantum light emission~\cite{Tan2020, Gil2022}. One of the most intriguing and technologically relevant aspects of $h$-BN is its tendency to form various stacking sequences, or \textit{polytypes}, due to the nearly degenerate interlayer energy landscape. These polytypes--notably AA$^\prime$ ($h$-BN), ABC ($r$-BN), AA ($e$-BN), and AB ($b$-BN)--differ only in how their layers are arranged along the out-of-plane direction, yet exhibit markedly distinct electronic, vibrational, and optical properties~\cite{Gil2022, Olovsson2022, Liu2003, Gilbert2019, Novotn2023}. The co-existence of these polytypes within single crystals leads to structural disorder, interfaces, and stacking faults, which can affect charge transport, phonon lifetimes, and emission characteristics~\cite{Ordin1998, Olovsson2022}.

Despite extensive research, the thermodynamic and vibrational stability of boron nitride (BN) polymorphs remains a topic of active investigation and debate. While cubic BN ($c$-BN) is the ground state under high pressure, recent high-level calculations and experiments suggest that $h$-BN is thermodynamically stabilized at ambient conditions through entropic contributions and many-body van der Waals (vdW) interactions~\cite{Cazorla2019, Nikaido2022}. In particular, the subtle energy differences between the low-energy $sp^2$ phases (such as $h$-BN and $r$-BN) are on the order of tens of meV per formula unit--a scale that demands theoretical methods beyond conventional DFT to resolve~\cite{Cazorla2019, Nikaido2022, Korona2019}. Including nonlocal correlation effects is thus essential for capturing the relative stability and interlayer coupling in these materials. Stacking order is a structural detail and a critical determinant of electronic structure and vibrational response. For instance, AA$^\prime$ stacking (common in $h$-BN) is predicted to be the lowest energy configuration among the layered forms~\cite{Constantinescu2013}. In contrast, ABC stacking ($r$-BN) yields distinct optical and electronic properties, including a redshifted conduction band minimum and richer IR/Raman activity~\cite{Liu2003, Olovsson2022, Novotn2023}. Additionally, $b$-BN and $e$-BN represent higher-symmetry and more idealized stackings (AB and AA, respectively), which, while less studied experimentally, serve as critical structural models to explore the influence of interlayer registry and symmetry on physical properties~\cite{Gilbert2019}.

Experimental studies using cathodoluminescence, Raman scattering, X-ray absorption, UV photoluminescence, and IR spectroscopy have revealed that polytypism dramatically influences the optoelectronic response of BN~\cite{Gil2022, Olovsson2022, Ordin1998, Iwański2024}. Notably, recent photoluminescence studies have succeeded in distinguishing BN polytypes through subtle spectral shifts and emission signatures~\cite{Iwański2024, Korona2023}. However, interpreting these data is complicated by the frequent co-existence of polytypes, nanoscale disorder, and temperature-induced transformations~\cite{Cazorla2019, Nikaido2022}. To this end, computational modeling is pivotal in deconvoluting these effects and guiding polytype-specific device design.

In this work, we present a comprehensive first-principles investigation of the vibrational and dielectric properties of four BN polymorphs--$e$-BN (AA), $h$-BN (AA$^\prime$), $r$-BN (ABC), and $b$-BN (AB). We use density functional perturbation theory (DFPT) with and without vdW corrections to examine the stacking-dependent evolution of phonon frequencies, Raman and infrared activities, and dielectric properties. Our results provide insight into the role of interlayer interactions, symmetry, and dispersion forces in determining the vibrational landscape of BN polymorphs and complement recent experimental efforts in distinguishing polytypes using vibrational spectroscopy. Understanding these intricate relationships is crucial for the fundamental science of 2D materials and for engineering BN-based optoelectronic and quantum devices with targeted performance. The findings presented here aim to bridge this gap and contribute to a more predictive understanding of polytype-dependent properties in layered BN.

\section{Computational Details}

We employed density functional theory (DFT) as implemented in the \textsc{Quantum ESPRESSO} (QE) package to investigate the structural, electronic, and vibrational properties of different BN phases. The calculations were initiated using the experimental unit cell, with explicitly defined lattice parameters in Cartesian coordinates. A plane-wave energy cutoff of 80~Ry was chosen to ensure numerical accuracy, following a systematic convergence analysis. The exchange-correlation interactions were treated using the Perdew-Burke-Ernzerhof (PBE) functional within the generalized gradient approximation (GGA). Brillouin zone sampling was performed using a $\Gamma$-centered $16 \times 16 \times 16$ Monkhorst-Pack $k$ point mesh to ensure accurate integration. Given the layered nature of BN, vdW corrections were incorporated using the DFT-D approach~\cite{Barone2009}. The convergence criteria were set to $1.0 \times 10^{-5}$ Ry for total energy and $1.0 \times$ $10^{-4} \mathrm{Ry} /$ Bohr for atomic forces. Projector-augmented wave (PAW) pseudopotentials were employed for electron-ion interactions, providing a reliable description of core-valence interactions while maintaining computational efficiency. The cohesive energy per atom, $E_c$, was calculated as
$
E_c=\frac{n_{\mathrm{B}} E_{\mathrm{B}}^{\text {atom }}+n_{\mathrm{N}} E_{\mathrm{N}}^{\text {atom }}-E_{\mathrm{BN}}}{N}
,$
where $E_{\mathrm{B}}^{\text {atom }}$ and $E_{\mathrm{N}}^{\text {atom }}$ are the total energies of isolated boron and nitrogen atoms, respectively; $n_{\mathrm{B}}$ and $n_{\mathrm{N}}$ are the numbers of B and N atoms in the BN unit cell; $E_{\mathrm{BN}}$ is the total energy of the BN polymorph; and $N=n_{\mathrm{B}}+n_{\mathrm{N}}$ is the total number of atoms. The atomic energies were computed in large cubic cells to eliminate spurious interactions.


\section{Results}
\subsection{Crystal structures for different polymorphs of BN}

\begin{figure}
    \centering
    \subfigure[\textbf{$e$-BN (AA)}]{
        \begin{minipage}[b]{0.225\textwidth}
            \centering
            \includegraphics[width=0.7\linewidth]{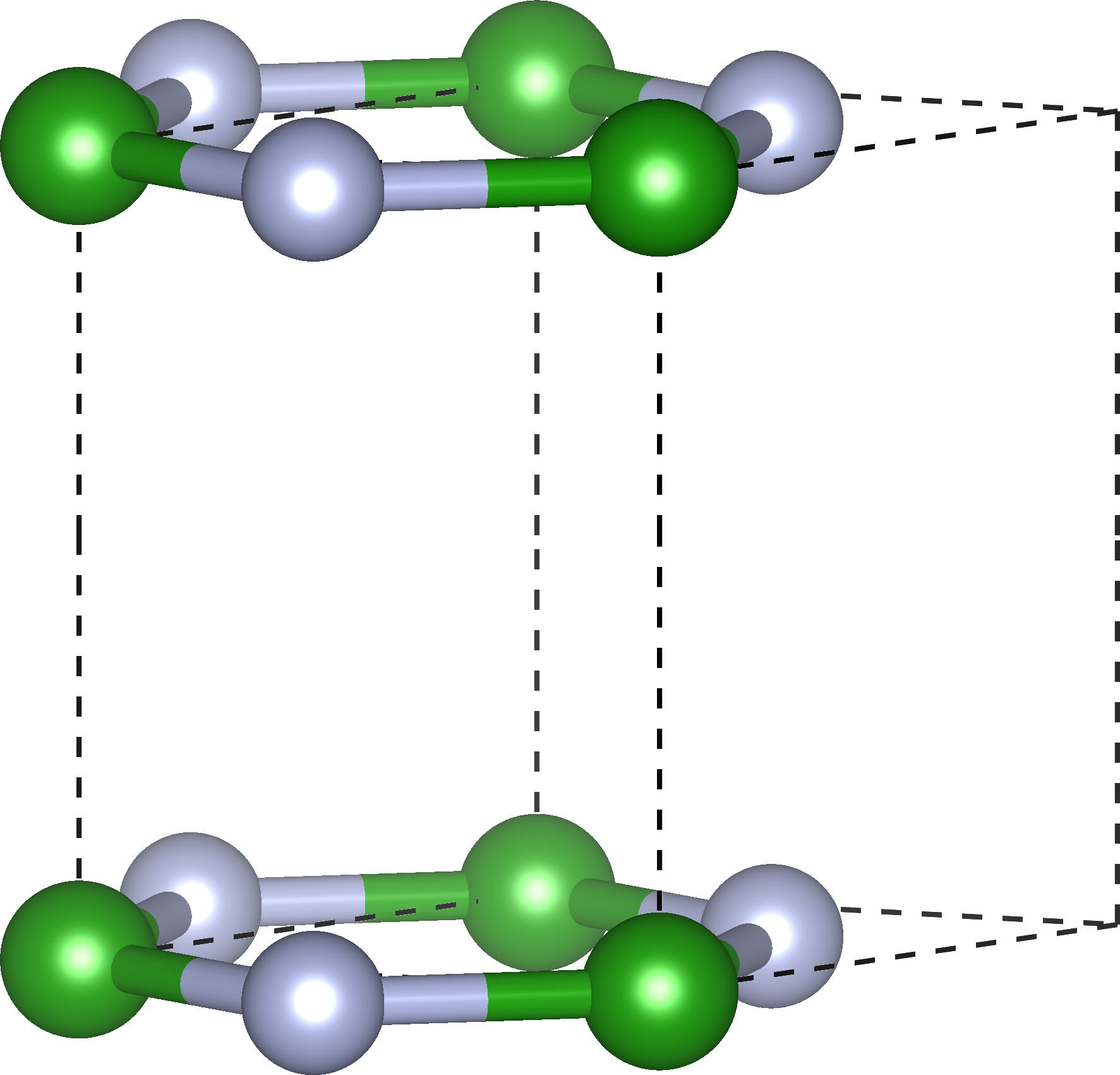}
        \end{minipage}
    }
    \subfigure[\textbf{$h$-BN (AA$^\prime$)}]{
        \begin{minipage}[b]{0.225\textwidth}
            \centering
            \includegraphics[width=0.9\linewidth]{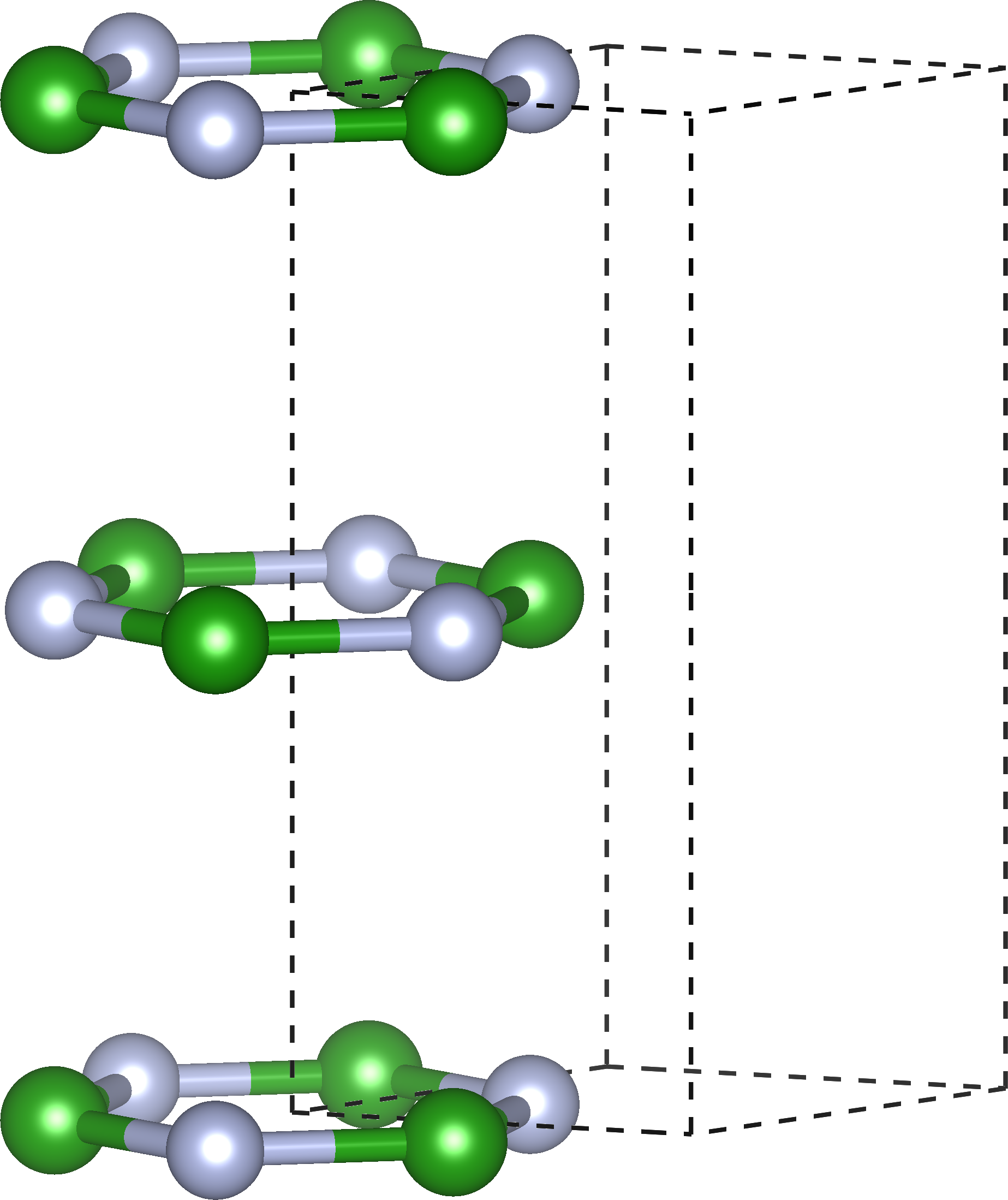}
        \end{minipage}
    } \\
    \subfigure[\textbf{$r$-BN (ABC)}]{
        \begin{minipage}[b]{0.225\textwidth}
            \centering
            \includegraphics[width=0.9\linewidth]{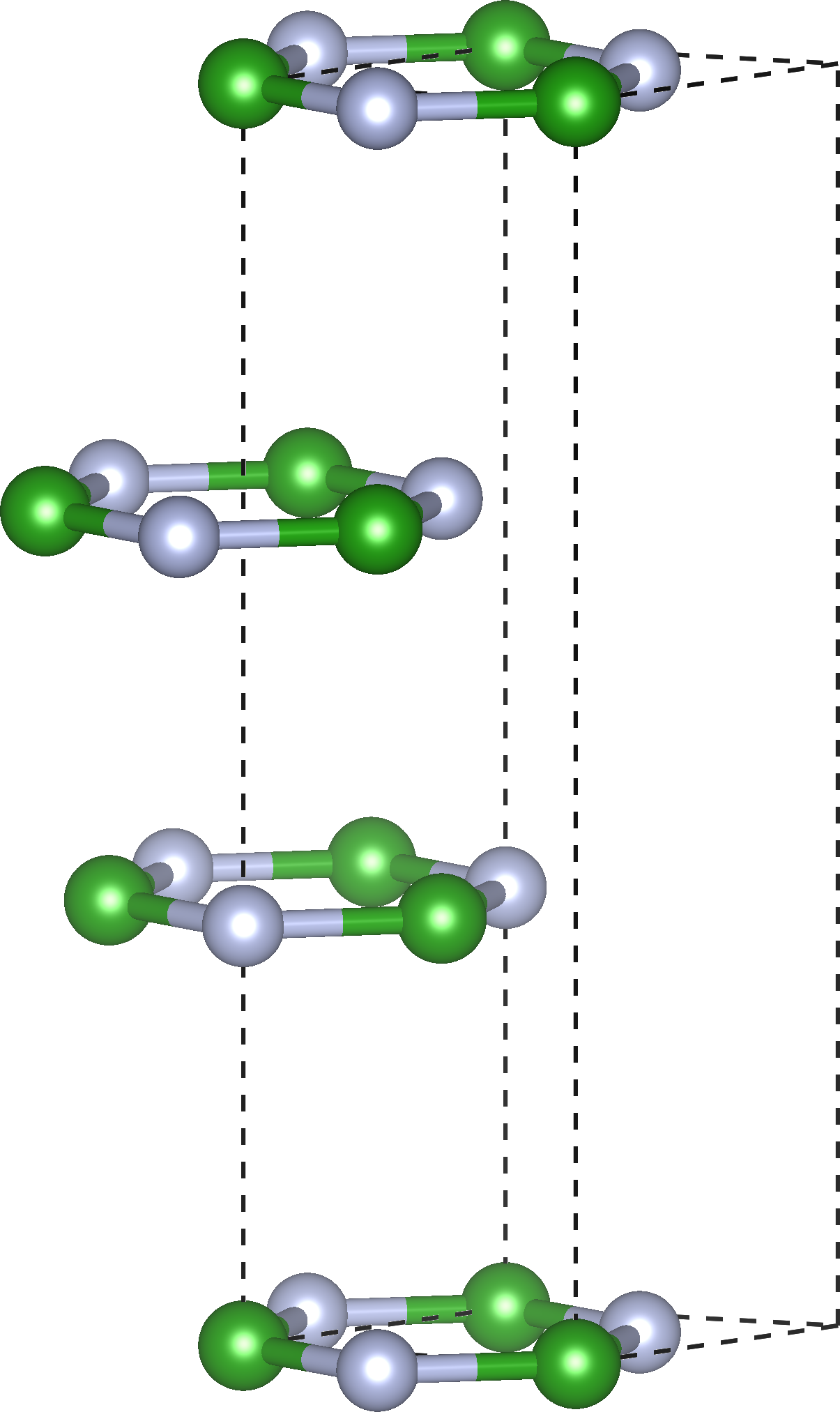}
        \end{minipage}
    }
    \subfigure[\textbf{$b$-BN (AB)}]{
        \begin{minipage}[b]{0.225\textwidth}
            \centering
            \includegraphics[width=0.9\linewidth]{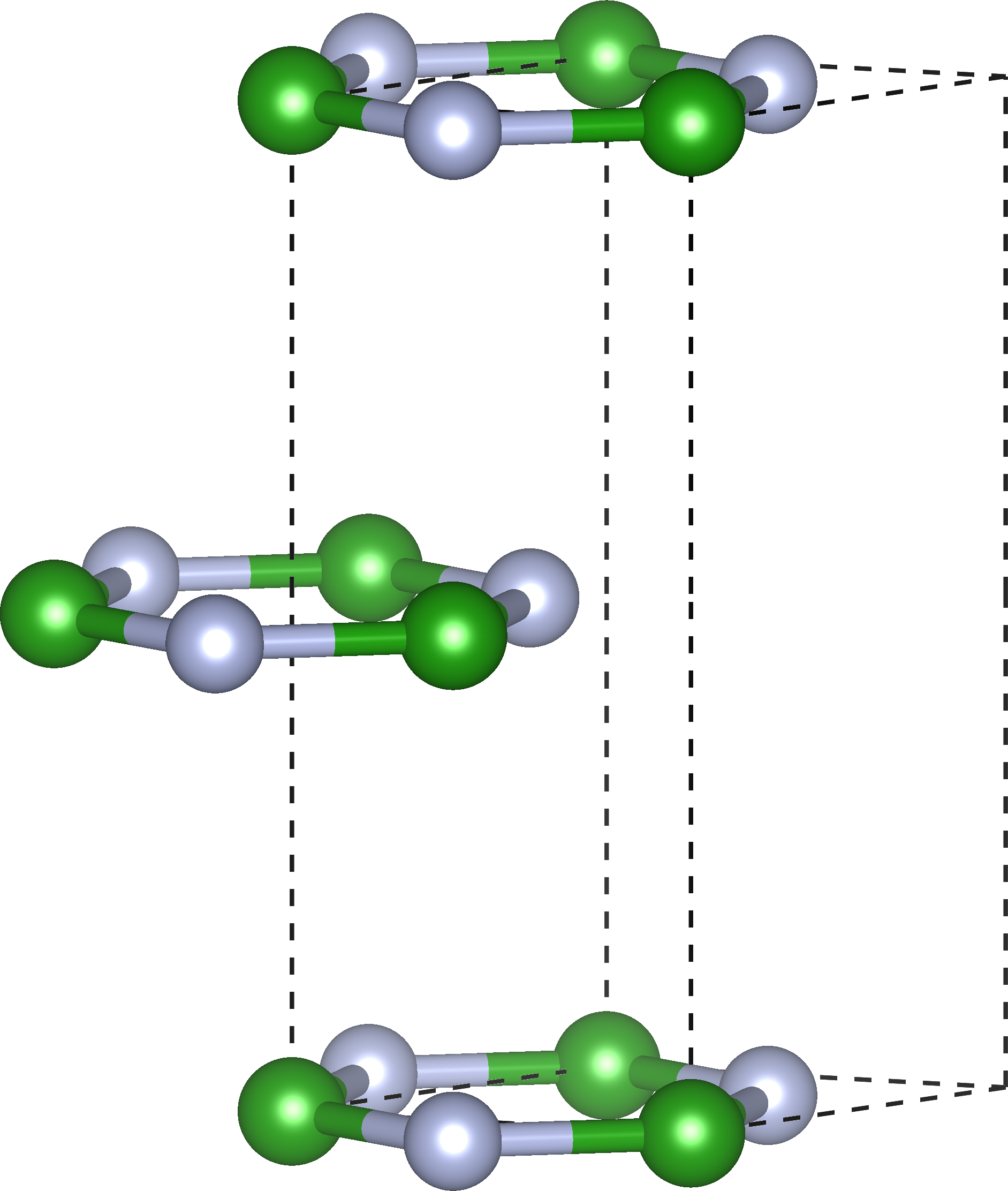}
        \end{minipage}
    }
   \caption{
        Crystal structures of BN polymorphs (green = B, gray = N):
        (a) $e$-BN [$P\bar{6}m2$ (187)],
        (b) $h$-BN [$P6_3/mmc$ (194)],
        (c) $r$-BN [$R\bar{3}m$ (166)], and
        (d) $b$-BN [$P6_3/mmc$ (194)].
    }
    \label{fig:bn_structures}
\end{figure}

The atomic structures of four layered BN polymorphs--$e$-BN (AA), $h$-BN (AA$^\prime$), $r$-BN (ABC), and $b$-BN (AB)--are illustrated in Figs.~\ref{fig:bn_structures}a--d. These polymorphs differ in stacking sequences, symmetry groups, and interlayer arrangements. Structural parameters, cohesive energies (\(E_c\)), and electronic band gaps (\(E_g\)) are summarized in Table~\ref{tab1}. The stacking order is a key factor controlling the interlayer spacing and stability of these $sp^2$-bonded systems.

$e$-BN (AA) crystallizes in the non-centrosymmetric space group $P\bar{6}m2$ (No.~187) and features direct AA stacking, where boron and nitrogen atoms are aligned vertically across layers (Fig.~\ref{fig:bn_structures}a). This configuration leads to weak interlayer bonding in the absence of vdW corrections. Inclusion of vdW interactions significantly contracts the \(c\)-axis from 5.06 to 3.38~\AA, as shown in Table~\ref{tab1}, improving agreement with values reported by Gil \textit{et al.}~\cite{Gil2022}. No experimental data exist for this stacking, but theoretical values are consistent across methods.

$h$-BN (AA$^\prime$) is the most commonly observed polymorph, adopting the $P6_3/mmc$ (No.~194) space group, with alternating B and N atoms stacked in a staggered bilayer configuration (Fig.~\ref{fig:bn_structures}b). The calculated lattice constants with vdW corrections (\(a = 2.512\)~\AA, \(c = 6.179\)~\AA) agree well with both theoretical and experimental values reported in Table~1 of Gil \textit{et al.}~\cite{Gil2022} and Ahmed \textit{et al.}~\cite{Ahmed2007}, where \(a = 2.478\)~\AA~and \(c = 6.354\)~\AA~are cited as experimental averages. This stacking is energetically favored and widely accepted as the most stable $sp^2$-BN structure under ambient conditions.

$r$-BN (ABC) belongs to the rhombohedral space group $R\bar{3}m$ (No.~166) and exhibits a three-layer ABC stacking sequence (Fig.~\ref{fig:bn_structures}c). The calculated interlayer distance (\(c = 9.17\)~\AA~with vdW) aligns reasonably with experimental reports (\(c = 9.679\)~\AA) in Table~1 of Gil \textit{et al.}~\cite{Gil2022}. Our calculations also show a slightly larger cohesive energy compared to $h$-BN (7.207 vs. 7.205~eV), consistent with the finding by Nikaido \textit{et al.}~\cite{Nikaido2022} that $r$-BN and $h$-BN are nearly degenerate in energy, although $h$-BN remains thermodynamically most stable at 0 K.

$b$-BN (AB) is a less commonly studied polymorph with $P6_3/mmc$ symmetry, characterized by a two-layer AB stacking, also called Bernal stacking (Fig.~\ref{fig:bn_structures}d). The optimized lattice parameters with vdW corrections (\(a = 2.511\)~\AA, \(c = 6.117\)~\AA) fall within the range of those reported for $sp^2$-BN systems in Gil \textit{et al.}~\cite{Gil2022}. The relative stability of $b$-BN is comparable to $r$-BN in our calculations, and it exhibits a direct band gap, suggesting unique electronic properties discussed in Sec.~\ref{sec:bandstructure}.

Across all polymorphs, we observe a strong sensitivity of the \(c\)-axis lattice parameter to vdW corrections, while the in-plane lattice constant \(a\) remains nearly invariant. This agrees with trends reported in both Gil \textit{et al.}~\cite{Gil2022} and Ahmed \textit{et al.}~\cite{Ahmed2007}, whose comprehensive tables include both theoretical and experimental lattice data. The cohesive energies calculated here suggest a delicate balance in stability among $sp^2$ BN polytypes, with $h$-BN slightly preferred, in line with diffusion Monte Carlo results by Nikaido \textit{et al.}~\cite{Nikaido2022}. These results highlight the essential role of interlayer stacking in modulating structural and energetic properties of layered BN and provide a consistent theoretical framework in support of experimental data. Overall, the structural polymorphism in $sp^2$-bonded BN results in a delicate balance between symmetry, interlayer stacking, and vdW interactions, all of which are crucial for understanding the phase stability and physical properties of BN-based materials.

\begin{table}
    \centering
    \caption{Comparison of structural parameters, cohesive energy (\(E_c\)), and electronic band gap (\(E_g\)) for layered BN polymorphs with and without vdW corrections.}
    \resizebox{\columnwidth}{!}{
    \begin{tabular}{lcccc}
        \hline \hline
        Method & \(a\) [\AA] & \(c\) [\AA] & \(E_c\) [eV] & \(E_g\) [eV] \\
        \hline \hline
        \multicolumn{5}{c}{\textbf{\emph{e}-BN (AA)}} \\
        \hline
        without vdW & 2.514 & 5.059 & 7.064 & 4.12 (indirect, K–$\Gamma$) \\
        with vdW    & 2.511 & 3.383 & 7.181 & 4.63 (indirect, K–$\Gamma$) \\
        literature  & 2.476~\cite{Gil2022} & 3.476~\cite{Gil2022} & – & – \\
        \hline
        \multicolumn{5}{c}{\textbf{\emph{h}-BN (AA$^\prime$)}} \\
        \hline
        without vdW & 2.515 & 9.136 & 7.065 & 4.25 (indirect, K–$\Gamma$) \\
        with vdW    & 2.512 & 6.179 & 7.205 & 4.10 (indirect, K–M) \\
        literature  & 2.478~\cite{Gil2022} & 6.354~\cite{Gil2022} & 7.055~\cite{Ahmed2007} & 4.25~\cite{Olovsson2022} \\
        \hline
        \multicolumn{5}{c}{\textbf{\emph{r}-BN (ABC)}} \\
        \hline
        without vdW & 2.515 & 13.743 & 7.065 & 4.24 (indirect, K–$\Gamma$) \\
        with vdW    & 2.511 & 9.168  & 7.207 & 3.94 (indirect, K–M) \\
        literature  & 2.476~\cite{Gil2022} & 9.679~\cite{Gil2022} & – & 4.21~\cite{Olovsson2022} \\
        \hline
        \multicolumn{5}{c}{\textbf{\emph{b}-BN (AB)}} \\
        \hline
        without vdW & 2.514 & 9.229 & 7.065 & 4.11 (indirect, K–$\Gamma$) \\
        with vdW    & 2.511 & 6.117 & 7.207 & 3.98 (direct, K–K) \\
        literature  & 2.477~\cite{Gil2022} & 6.319~\cite{Gil2022} & – & – \\
        \hline \hline
    \end{tabular}
    }
    \label{tab1}
\end{table}


\subsection{Electronic Band Structure}
\label{sec:bandstructure}

\begin{figure*}[ht]
    \centering
    \subfigure[$e$-BN]{\includegraphics[width=0.45\textwidth]{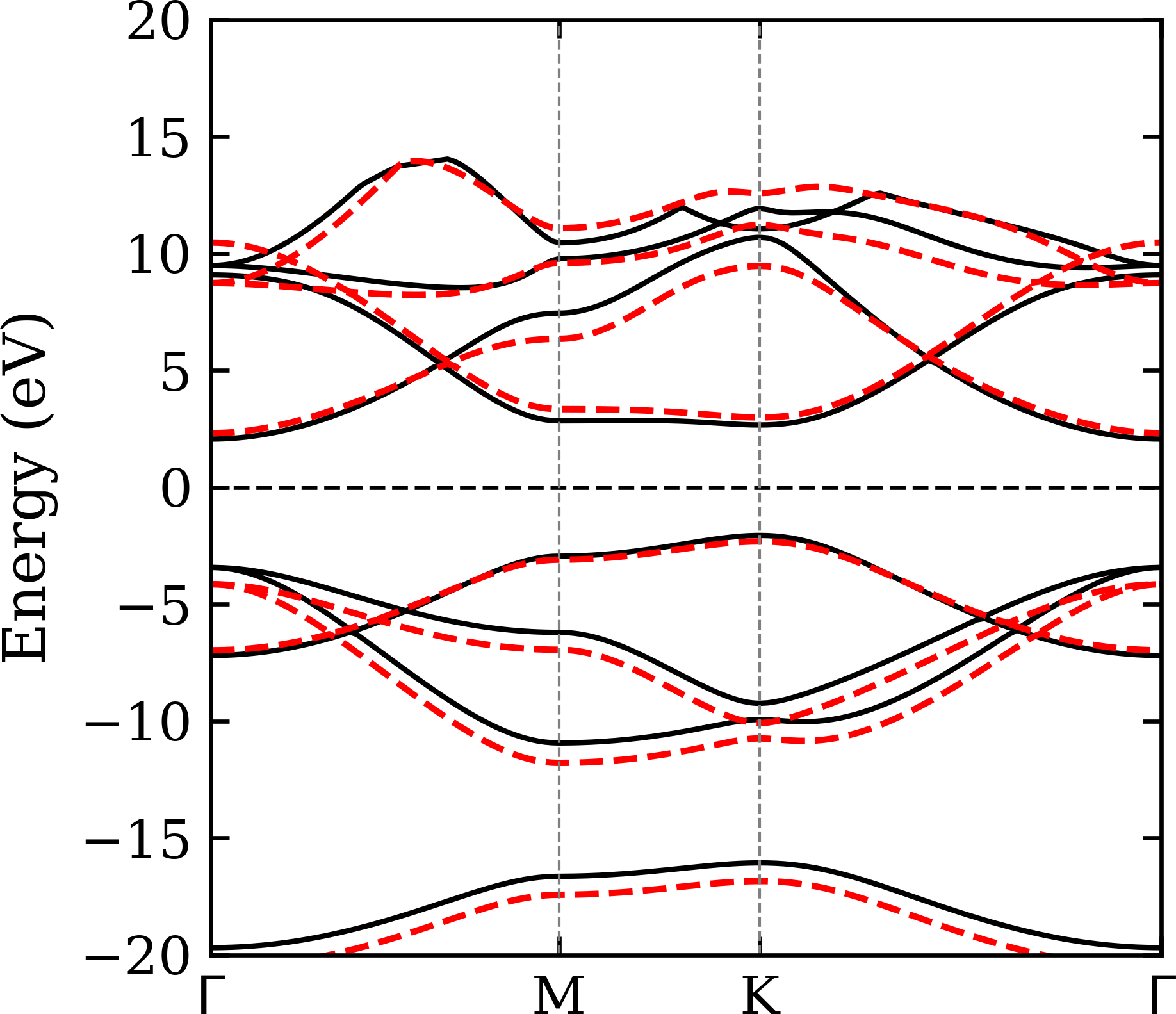}}
    \subfigure[$h$-BN]{\includegraphics[width=0.45\textwidth]{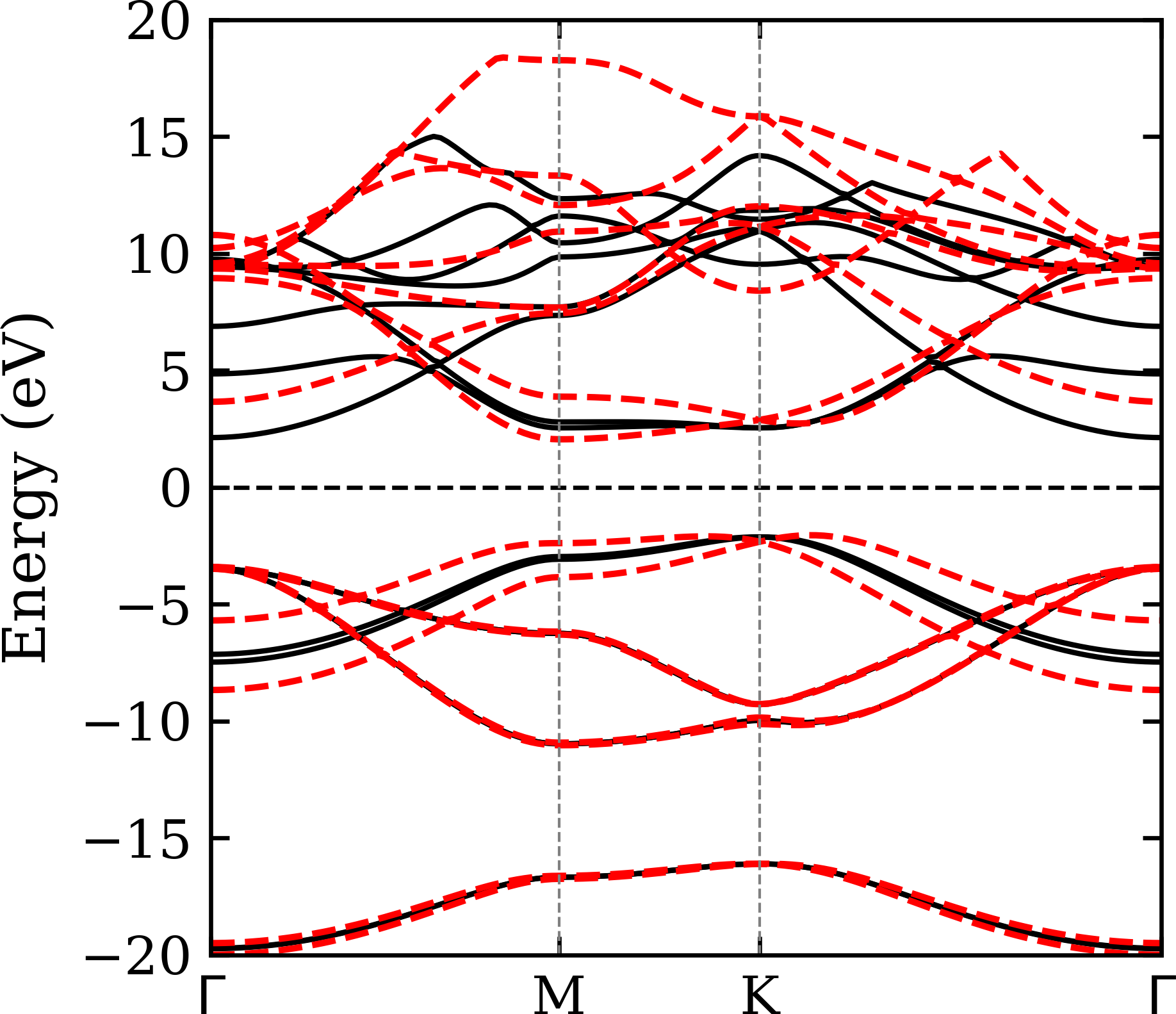}} \\
    \subfigure[$r$-BN]{\includegraphics[width=0.45\textwidth]{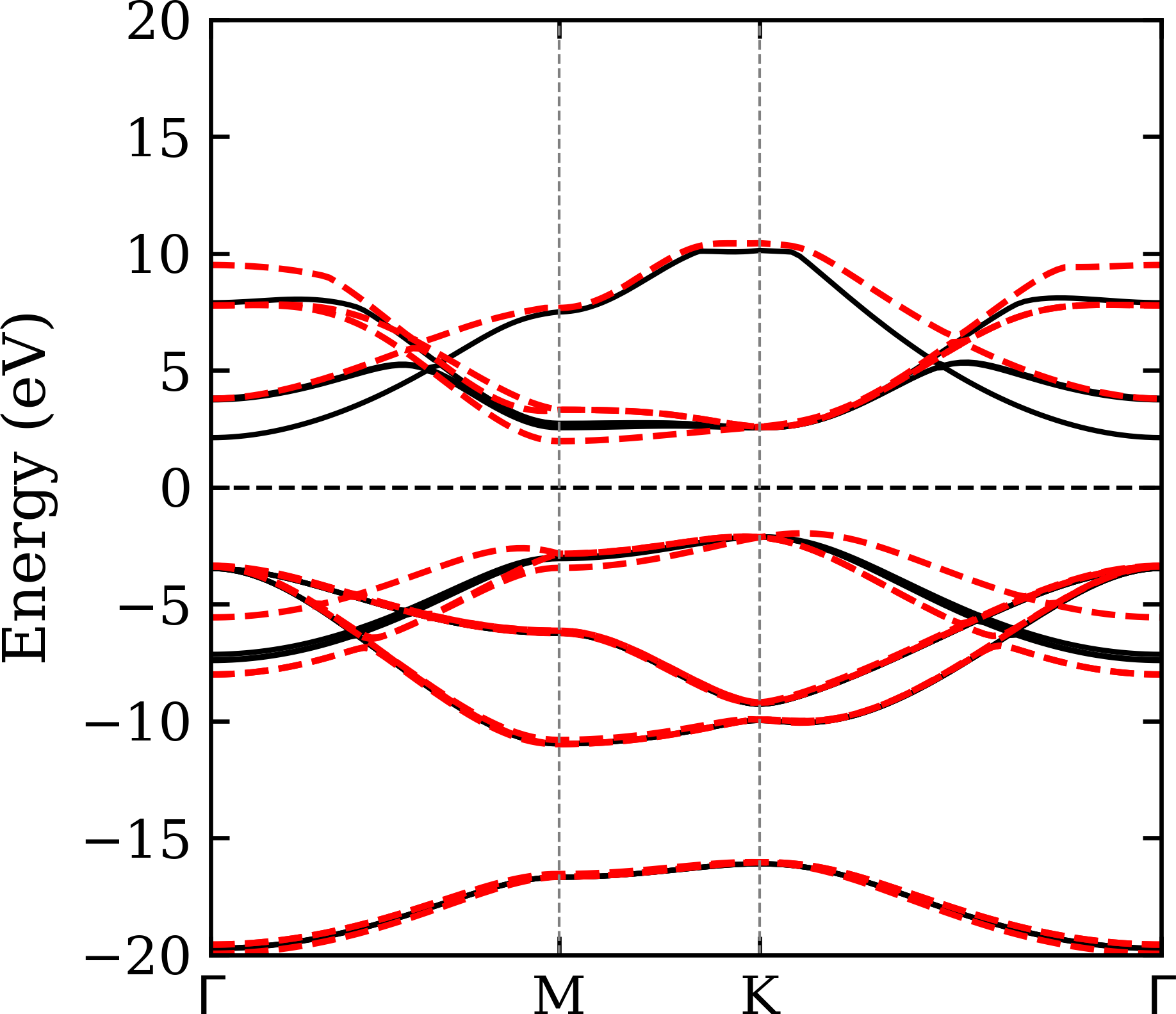}}
    \subfigure[$b$-BN]{\includegraphics[width=0.45\textwidth]{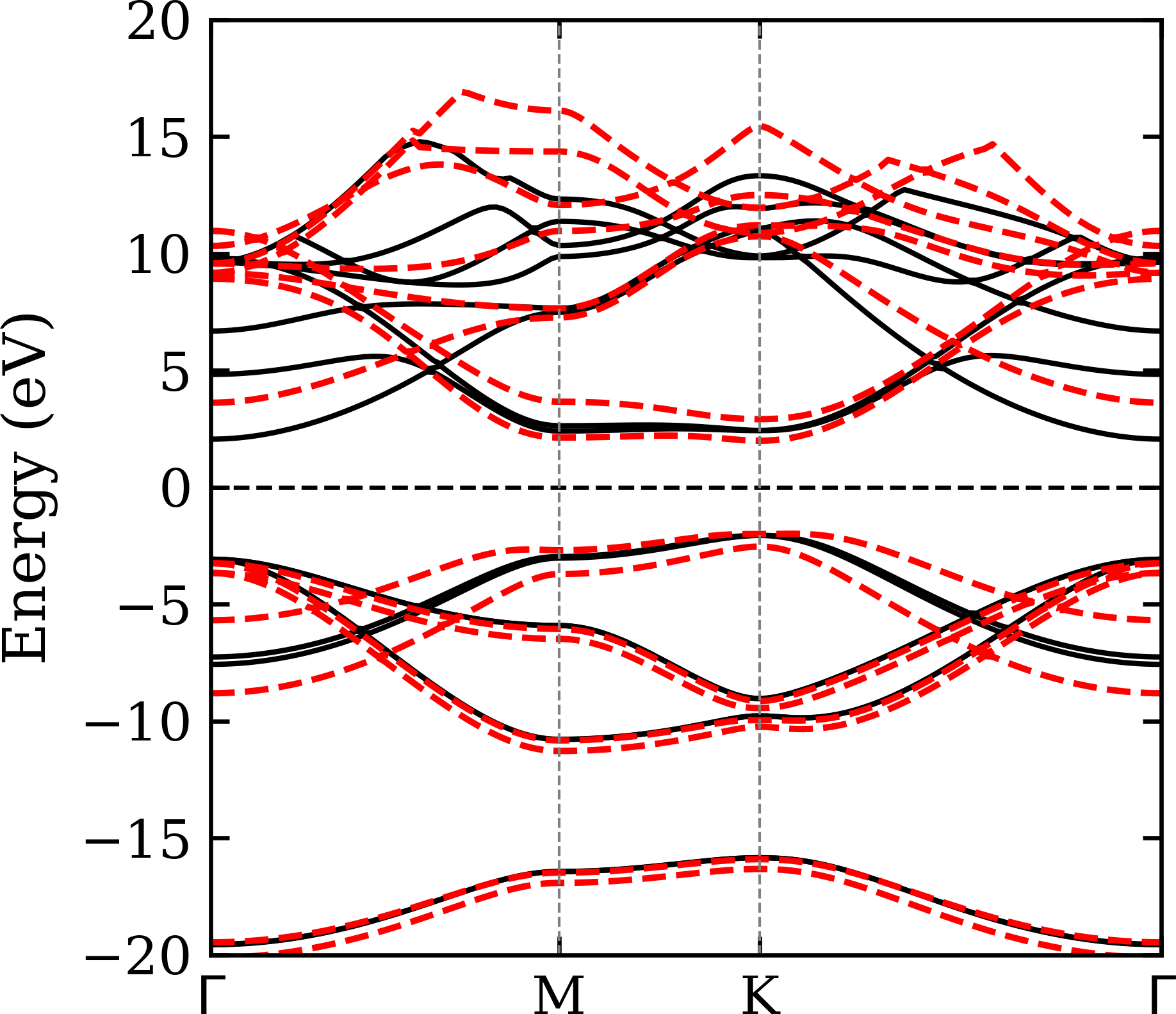}}
    \caption{Electronic band structures of different BN polymorphs with and without vdW corrections. Black solid lines indicate results without vdW, and red dashed lines include vdW corrections.}
    \label{fig:bands}
\end{figure*}

The electronic band structures of the four studied BN polymorphs were calculated with and without the vdW correction. The results are shown in Fig.~\ref{fig:bands}, while a summary of the band gaps' details is presented in Table~\ref{tab1}. All polymorphs exhibit wide band gaps in the range of 3.94 to 4.63~eV, consistent with the semiconducting nature of $sp^2$-bonded BN. The inclusion of vdW interactions significantly affects the out-of-plane lattice constants and, consequently, the electronic band structure, underscoring the importance of capturing weak interlayer forces in layered materials. Across the studied BN polymorphs, the electronic band structure reveals predominantly indirect band gaps, except for $b$-BN, which exhibits a direct transition at the K point. Specifically, $e$-BN, $h$-BN, and $r$-BN show indirect band gaps ranging from 3.94 to 4.63~eV, typically between the K and $\Gamma$ or K and M points. Among them, $e$-BN has the widest gap (4.63~eV), while $r$-BN has the narrowest (3.94~eV), highlighting the influence of stacking geometry and interlayer coupling. The reduced gap in $r$-BN, a rhombohedral phase, can be attributed to enhanced interlayer orbital overlap due to its three-layer ABC stacking, which modifies the conduction band minimum. In contrast, $b$-BN exhibits a direct band gap (K--K), a property desirable for optoelectronic applications relying on vertical transitions. This distinction underscores the sensitivity of the electronic structure to interlayer interactions and symmetry, and emphasizes the potential of stacking-engineered BN for tailored band gap applications.

Comparison with earlier work by Ahmed \textit{et al.}~\cite{Ahmed2007} shows qualitative agreement in the trends of band gap variations across BN polymorphs. Their full-potential LAPW calculations using the Engel--Vosko GGA functional yielded a gap of 4.18~eV for $h$-BN and 4.21~eV for $r$-BN, closely matching our PBE+vdW results. However, Ahmed \textit{et al.} emphasize the importance of using GGA-EV to better align with experimental values, suggesting a pathway for further refinement. Additional insight is provided by Olovsson and Magnuson~\cite{Olovsson2022}, who investigated $h$-, $r$-, and turbostratic BN using X-ray absorption near-edge structure (XANES) spectroscopy. Their DFT+core-hole simulations revealed distinctive $\pi^*$ and $\sigma^*$ features sensitive to stacking order. The observed shift of the $\pi^*$ onset in $r$-BN relative to $h$-BN correlates with our finding of a narrower band gap in the rhombohedral structure. Moreover, their turbostratic BN models exhibited an average band gap of approximately 3.86~eV, further supporting the notion that stacking disorder can be exploited to engineer BN’s electronic properties.

Lastly, our results show that interlayer stacking and vdW interactions are critical determinants of the electronic properties of BN polymorphs. The transition between indirect and direct band gaps, combined with variations in gap magnitude, offers opportunities for targeted design of BN-based materials in optoelectronics, UV photonics, and quantum applications.


\subsection{Phonon Frequencies}

\begin{table*}[ht]
\centering
\caption{
Phonon frequencies (in cm$^{-1}$) at the $\Gamma$ point for each BN polymorph, calculated using DFPT with and without vdW corrections. Each entry lists the frequency shift from no-vdW to vdW, followed by the irreducible representation and mode activity: infrared (I), Raman (R), both (I+R), or silent (S). Only optical phonons are included, with vdW-corrected values shown second in each pair.
}
\label{tab:phonon}
\begin{tabular}{cc cc cc cc}
\toprule
\multicolumn{2}{c}{$e$-BN} &
\multicolumn{2}{c}{$h$-BN} &
\multicolumn{2}{c}{$r$-BN} &
\multicolumn{2}{c}{$b$-BN} \\
\cmidrule(lr){1-2} \cmidrule(lr){3-4} \cmidrule(lr){5-6} \cmidrule(lr){7-8}
Freq. (no$\rightarrow$vdW) & Mode & Freq. (no$\rightarrow$vdW) & Mode & Freq. (no$\rightarrow$vdW) & Mode & Freq. (no$\rightarrow$vdW) & Mode \\
\midrule
783 $\rightarrow$ 757.2     & $A_2''$ (I)          & 0 $\rightarrow$ 39.6        & $E_{2g}$ (R)       & 0 $\rightarrow$ 37.3       & $E$ (I+R)         & 0 $\rightarrow$ 48.5      & $E'$ (I+R) \\
1343.5 $\rightarrow$ 1352.5 & $E'$ (I+R)           & 52.9 $\rightarrow$ 182.8    & $B_{1g}$ (S)       & 0 $\rightarrow$ 39.3       & $E$ (I+R)         & 55.2 $\rightarrow$ 180.0   & $A_2''$ (I) \\
                            &                      & 781.1 $\rightarrow$ 723     & $A_{2u}$ (I)       & 0 $\rightarrow$ 150.2      & $A_1$ (I+R)       & 781.6 $\rightarrow$ 732.0  & $A_2''$ (I) \\
                            &                      & 803.0 $\rightarrow$ 792.1   & $B_{1g}$ (S)       & 49.2 $\rightarrow$ 159.2   & $A_1$ (I+R)       & 803.0 $\rightarrow$ 793.9  & $A_2''$ (I) \\
                            &                      & 1341.6 $\rightarrow$ 1349.3 & $E_{1u}$ (I)       & 778.5 $\rightarrow$ 730.6  & $A_1$ (I+R)       & 1343.6 $\rightarrow$ 1351.9 & $E'$ (I+R) \\
                            &                      & 1341.6 $\rightarrow$ 1350.3 & $E_{2g}$ (R)       & 799.2 $\rightarrow$ 795.7  & $A_1$ (I+R)       & 1343.6 $\rightarrow$ 1358.7 & $E'$ (I+R) \\
                            &                      &                             &                    & 801.4 $\rightarrow$ 797.8  & $A_1$ (I+R)       &                             &            \\
                            &                      &                             &                    & 1343.2 $\rightarrow$ 1353.1 & $E$ (I+R)         &                             &            \\
                            &                      &                             &                    & 1343.2 $\rightarrow$ 1356.4 & $E$ (I+R)         &                             &            \\
                            &                      &                             &                    & 1343.3 $\rightarrow$ 1356.4 & $E$ (I+R)         &                             &            \\
\bottomrule
\end{tabular}
\end{table*}

The vibrational properties of layered BN polymorphs are susceptible to weak interlayer forces and stacking configurations. To accurately capture these effects, we computed phonon frequencies at the $\Gamma$ point of the Brillouin zone using DFPT with and without vdW corrections, as implemented in QE. The refined phonon spectra--including full irreducible representation labeling and explicit identification of silent (S), infrared (I), and Raman (R) active modes--are summarized in Table~\ref{tab:phonon}. All optical phonons are included, providing a comprehensive mode-by-mode comparison across BN polymorphs. Including vdW interactions significantly impacts phonon frequencies, particularly in polymorphs with strong interlayer coupling such as $h$-BN and $r$-BN. Consistent with prior studies~\cite{Gil2022,Nikaido2022}, vdW corrections soften out-of-plane optical modes and bring theoretical spectra closer to experimental IR and Raman observations. The detailed mode analysis reveals prominent shifts upon vdW inclusion, especially for low-frequency modes below 200~cm$^{-1}$ and high-frequency optical branches near 1350~cm$^{-1}$. For example, in $h$-BN, the out-of-plane infrared-active $A_{2u}$ mode shifts from 781.1~cm$^{-1}$ (no vdW) to 723~cm$^{-1}$ (with vdW), while the silent $B_{1g}$ mode also softens notably. Similarly, $r$-BN exhibits rich vibrational behavior with multiple modes showing dual IR and Raman activity (labeled as I+R), especially between 730–800~cm$^{-1}$, where vdW corrections shift the $A_1$ modes by 20–40~cm$^{-1}$. In $e$-BN, the phonon spectrum reflects the high symmetry and absence of staggered stacking. We observe fewer distinct branches and minimal splitting between optical modes. The dominant IR-active $A^{\prime\prime}_2$ and $E^{\prime}$ modes remain near 780~cm$^{-1}$ and 1343~cm$^{-1}$, respectively, with vdW corrections introducing only moderate shifts. Notably, silent modes, such as those transforming as $B_{1g}$ in analogous systems, are absent here due to symmetry constraints. Conversely, $b$-BN shows a broader vibrational landscape. Low-frequency IR-active modes shift from ~55~cm$^{-1}$ (no vdW) to ~180~cm$^{-1}$ (with vdW), reflecting stronger interlayer coupling. Intermediate-frequency $A_1$ and $E^{\prime}$ modes also emerge around 730–800~cm$^{-1}$, with rich IR and Raman activity reflecting the lower symmetry of the AB stack. The high-frequency Raman-active $E^{\prime}$ modes near 1350~cm$^{-1}$ remain prominent and shift modestly upon vdW inclusion.

These results directly connect to the subsequent analysis of Raman and IR intensities (Figs.~\ref{fig:raman} and~\ref{fig:ir}), where stacking-dependent features--such as the rich mid-frequency IR activity of $r$-BN and the sharper Raman peaks of $e$-BN--mirror the phonon characteristics detailed here. Overall, this detailed vibrational analysis highlights the critical role of vdW interactions and stacking order in shaping the phonon spectra of layered BN polymorphs. The combination of full irreducible representation labeling and activity classification provides a robust framework for both theoretical interpretation and experimental verification of BN polytypes.


\subsection{Raman and Infrared Intensities}

\begin{figure}
    \centering
    \includegraphics[width=1.0\linewidth]{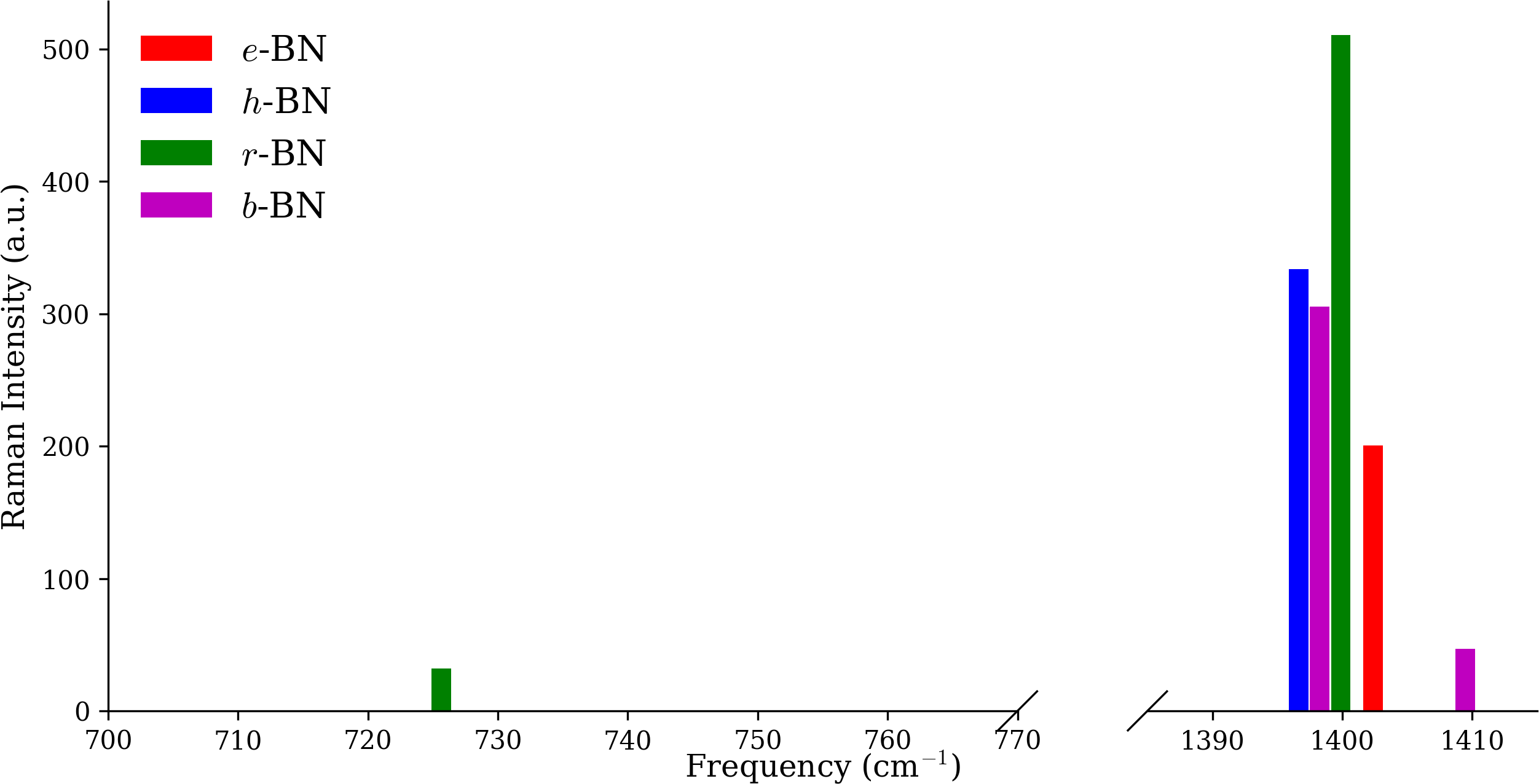}
    \caption{Raman intensity at the $\Gamma$ point for different BN polymorphs, computed using LDA-based ONCV pseudopotentials with vdW corrections.}
    \label{fig:raman}
\end{figure}

\begin{figure}
    \centering
    \includegraphics[width=1.0\linewidth]{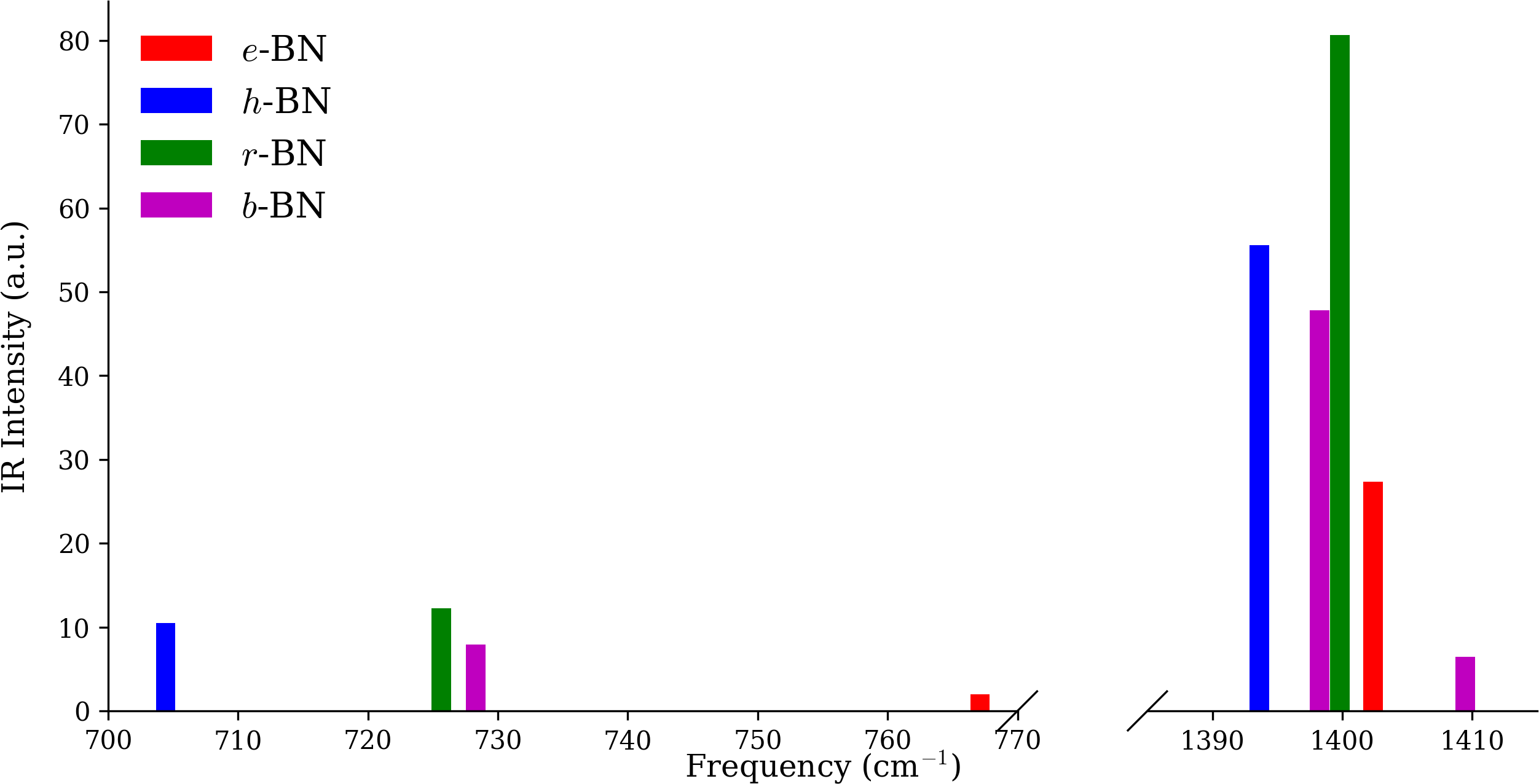}
    \caption{Infrared intensity at the $\Gamma$ point for different BN polymorphs, computed using LDA-based ONCV pseudopotentials with vdW corrections.}
    \label{fig:ir}
\end{figure}

To obtain accurate infrared (IR) and Raman intensities, pseudopotentials compatible with the linear and nonlinear response formalism in QE must be used. In particular, norm-conserving (NC) pseudopotentials based on the local density approximation (LDA) are required for reliable Raman intensity calculations due to the need for second-order response properties such as polarizability derivatives. To this end, we reoptimized the lattice parameters and atomic positions of each BN polymorph using optimized NC Vanderbilt pseudopotentials (ONCVPs) from the \textsc{PseudoDojo} project~\cite{vanSetten2018}, with all calculations performed within LDA. These LDA-ONCVP-based results, including vdW corrections, for phonon frequencies listed in the Supplementary Table~I agree reasonably well with the phonon frequencies shown in Table~\ref{tab:phonon}.

Figures~\ref{fig:raman} and~\ref{fig:ir} show the Raman and IR intensities computed at the $\Gamma$ point for the four studied BN polymorphs. The stacking-dependent optical response is clearly observed, in direct correspondence with the phonon mode distribution in Table~\ref{tab:phonon} and Supplementary Table~I. In the Raman spectra (Fig.~\ref{fig:raman}), all polymorphs exhibit a strong high-frequency peak near 1350~cm$^{-1}$, corresponding to in-plane $E_{2g}$ or $E^{\prime}$ modes of the $sp^2$-bonded BN lattice. The Raman activity of these modes varies significantly with stacking: it is strongest in $r$-BN and $h$-BN, consistent with enhanced interlayer polarizability anisotropy. These trends are reinforced by the large Raman intensities calculated for $r$-BN modes near 1398–1409~cm$^{-1}$ (see Supplementary Table~II). The IR spectra (Fig.~\ref{fig:ir}) similarly reveal distinct stacking fingerprints. $r$-BN exhibits rich IR activity between 730–795~cm$^{-1}$, arising from out-of-plane transverse optical (TO) modes sensitive to interlayer coupling. In contrast, $e$-BN shows a more limited IR response, reflecting its higher symmetry and fewer IR-active modes. The $b$-BN phase exhibits distributed IR and Raman activity in the mid-frequency region (600–800~cm$^{-1}$), likely due to complex interlayer interactions in its AB stacking.

These spectroscopic trends confirm the critical role of stacking configuration in shaping the optical activity of phonons. The complete symmetry labeling, IR/Raman classification, and the intensities provided in the Supplementary Information offer fingerprints for differentiating BN polytypes. Moreover, this analysis underscores the importance of using consistent pseudopotentials and exchange-correlation functionals when modeling spectroscopic properties of layered materials.


\section{Summary}

In summary, this study presents a comprehensive investigation of the structural, vibrational, and optical properties of layered BN polymorphs, emphasizing the essential role of vdW interactions and stacking configuration. By comparing results with and without vdW corrections, we demonstrate that dispersion forces are necessary for accurately describing the energetic, structural, and dynamical properties of $e$-BN, $h$-BN, $r$-BN, and $b$-BN. Including vdW corrections significantly improves agreement with experimental lattice constants and electronic band gaps and resolves stacking-dependent vibrational features. Our phonon analysis reveals strong sensitivity of low-frequency and out-of-plane optical modes to interlayer forces, particularly in $h$-BN and $r$-BN. All polymorphs are mechanically stable, reflected by the absence of imaginary phonon modes. Raman and IR intensities exhibit clear trends that reflect stacking order. For example, $r$-BN displays rich mid-frequency IR activity and strong Raman features due to LO–TO splitting, while $e$-BN exhibits sharp and well-isolated modes due to its high symmetry. These spectral fingerprints are experimentally relevant and enable precise polytype identification. It should be noted that slight numerical discrepancies may arise between different data sets due to variations in computational settings--such as the use of PAW versus ONCV pseudopotentials or the choice of exchange-correlation functional (e.g., PBE vs. LDA)--which influence the absolute values of phonon frequencies. Overall, this work highlights the need for including vdW interactions in first-principles modeling of layered materials and establishes a consistent computational framework using validated pseudopotentials to interpret and predict the vibrational and spectroscopic behavior of BN polytypes. Our findings aid in designing and identifying BN-based materials for advanced optoelectronic and quantum applications.

\bibliography{v5}

\clearpage
\onecolumngrid
\setcounter{table}{0}

\section*{Supplementary Information}

In this Supplementary Information, we provide the full set of phonon frequencies and corresponding Raman and IR intensities for the four BN polymorphs: $e$-BN (AA stacking), $h$-BN (AA$^\prime$ stacking), $r$-BN (ABC stacking), and $b$-BN (AB stacking). These calculations were performed using ONCVPs within the LDA approximation, which is necessary for accurate Raman and IR intensity predictions in QE. All frequencies are reported at the $\Gamma$ point with the vdW correction included. Symmetry assignments are also provided according to the relevant point group for each polymorph.

\begin{table*}[ht]
\centering
\caption{
Optical phonon frequencies (in cm$^{-1}$) at the $\Gamma$ point for each BN polymorph, including vdW corrections. Each mode is labeled with its irreducible representation and activity type: infrared (I), Raman (R), both (I+R), or silent (S).
}
\label{tab:freq_modes}
\begin{tabular}{cc cc cc cc}
\toprule
\multicolumn{2}{c}{$e$-BN} &
\multicolumn{2}{c}{$h$-BN} &
\multicolumn{2}{c}{$r$-BN} &
\multicolumn{2}{c}{$b$-BN} \\
\cmidrule(lr){1-2} \cmidrule(lr){3-4} \cmidrule(lr){5-6} \cmidrule(lr){7-8}
Frequency & Mode & Frequency & Mode & Frequency & Mode & Frequency & Mode \\
\midrule
767.2  & $A_2''$ (I)        & 84.6    & $E_{2g}$ (R)       & 83.1    & $E$ (I+R)      & 93.2    & $E'$ (I+R) \\
1402.4 & $E'$ (I+R)         & 241.0   & $B_{1g}$ (S)         & 83.2    & $E$ (I+R)      & 245.9   & $A_2''$ (I) \\
       &                    & 704.4   & $A_{2u}$ (I)       & 214.6   & $A_1$ (I+R)      & 728.3   & $A_2''$ (I) \\
       &                    & 816.8   & $B_{1g}$ (S)       & 214.6   & $A_1$ (I+R)      & 821.4   & $A_2''$ (I) \\
       &                    & 1393.6  & $E_{1u}$ (I)       & 725.7   & $A_1$ (I+R)      & 1398.3  & $E'$ (I+R) \\
       &                    & 1396.7  & $E_{2g}$ (R)       & 824.3   & $A_1$ (I+R)      & 1409.4  & $E'$ (I+R) \\
       &                    &         &                    & 824.4   & $A_1$ (I+R)      &         &            \\
       &                    &         &                    & 1399.9  & $E$ (I+R)   &         &            \\
       &                    &         &                    & 1405.4  & $E$ (I+R)      &         &            \\
       &                    &         &                    & 1405.4  & $E$ (I+R)      &         &            \\
\bottomrule
\end{tabular}
\end{table*}

\begin{table*}[ht]
\centering
\caption{
Calculated Raman and IR intensities for each BN polymorph at the $\Gamma$ point. Intensities are reported in arbitrary units. Frequencies are in cm$^{-1}$.
}
\label{tab:intensities}
\begin{tabular}{ccc ccc ccc ccc}
\toprule
\multicolumn{3}{c}{$e$-BN} &
\multicolumn{3}{c}{$h$-BN} &
\multicolumn{3}{c}{$r$-BN} &
\multicolumn{3}{c}{$b$-BN} \\
\cmidrule(lr){1-3} \cmidrule(lr){4-6} \cmidrule(lr){7-9} \cmidrule(lr){10-12}
Freq. & IR & Raman & Freq. & IR & Raman & Freq. & IR & Raman & Freq. & IR & Raman \\
\midrule
767.1    & 1.95    &         & 84.2    &        & 0.32    & 83.1    & 0       & 0         & 93.2    & 0.0035  & 0.16    \\
1402.4   & 27.32   & 200.54  & 704.4   & 10.5   &         & 83.2    & 0       & 0         & 245.9   & 0.0696  &         \\
         &         &         & 1393.6  & 55.56  &         & 214.6   & 0       & 0         & 728.3   & 7.95    &         \\
         &         &         & 1396.7  &        & 333.68  & 214.6   & 0       & 0         & 821.4   & 0.0003  &         \\
         &         &         &         &        &         & 725.7   & 12.27   & 32.17     & 1398.3  & 47.79   & 305.55  \\
         &         &         &         &        &         & 824.3   & 0       & 0         & 1409.5  & 6.46    & 46.69   \\
         &         &         &         &        &         & 824.4   & 0       & 0         &         &         &         \\
         &         &         &         &        &         & 1399.9  & 80.66   & 510.66    &         &         &         \\
         &         &         &         &        &         & 1405.4  & 0       & 0         &         &         &         \\
         &         &         &         &        &         & 1405.4  & 0       & 0         &         &         &         \\
\bottomrule
\end{tabular}
\end{table*}

\end{document}